\documentclass{article}
\usepackage{epsfig}
\begin{document}
\title{
{\normalsize \hfill SPIN-2003/26}\\
\vspace{-1.4cm}
{\normalsize \hfill ITP-UU-03/42}\\
${}$\\ 
${}$\\ 
Space-time foam in 2D and the sum over topologies%
\thanks{Presented by W.W. at the Workshop on Random Geometry, Krakow, May 15-17, 2003.}%
}
\author{R. Loll and W.Westra\\
${}$\\ 
{ Institute for Theoretical Physics, Utrecht University,}\\
{ Leuvenlaan 4, NL-3584 CE Utrecht, The Netherlands}
}
\maketitle
\begin{abstract}
It is well-known that the sum over topologies in quantum gravity is
ill-defined, due to a super-exponential growth of the number of
geometries as a function of the space-time volume, leading to
a badly divergent gravitational path integral. Not even in dimension 2, 
where a non-perturbative quantum gravity theory can be 
constructed explicitly from a (regularized) path integral, has this
problem found a satisfactory solution. -- In the present work,
we extend a previous 2d Lorentzian path integral,
regulated in terms of Lorentzian random triangulations,
to include space-times with an arbitrary number of handles.
We show that after the
imposition of physically motivated causality constraints, the
combined sum over geometries and topologies is well-defined and
possesses a continuum limit which yields a concrete model of 
space-time foam in two dimensions.
\end{abstract}

\section{Quantum gravity and the sum over topologies}

Many attempts of constructing a non-perturbative path integral
for gravity start from the premise that this should also
contain a sum over space-time topologies, formally written as
\begin{equation}
Z=\sum_{topol.} \int {\cal D}g_{\mu\nu}\ {\rm e}^{i S[g_{\mu\nu}]},
\label{pi}
\end{equation}
with the action
\begin{equation}
S=\int d^4x \sqrt{|\det g|} (\kappa R-\lambda),
\label{action}
\end{equation}
where each term in the sum (\ref{pi}) is given by the functional integral 
over equivalence classes $[g_{\mu\nu}]$ of metrics on
a space-time of a particular
topology. This assertion is usually followed by immediately
dropping
the sum again, since no way can be found to enumerate the
different topologies, let alone perform the sum explicitly. 

Needless to say that this state of affairs is highly unsatisfactory.
Whether or not a sum over topologies should be included is
connected to the nature of the fundamental degrees of freedom
governing quantum gravity at the very shortest scale, about
which little is known. Topological excitations seem a
natural enough candidate, and pictures of a multiply-connected
space-time foam\footnote{See \cite{garay} for a review and bibliography.} 
may be suggestive to the imagination,
but there is so far little direct or indirect evidence that such
structures are realized in nature. 

Is there then anything we can say about the issue of topology
change\footnote{Performing a sum over (space-time) topologies
in a path integral with fixed initial and final boundary conditions
implies configurations whose spatial topology changes in time.
For reviews of the issue of topology change in gravity, see
\cite{horo,dowker}.},
in the absence of a full-fledged non-perturbative theory
of quantum gravity? If we managed to make sense,
mathematically and physically, of the sum over topologies, 
how would the final theory be affected by the inclusion? Any 
theory predicting finite probabilities for {\it macroscopic} topology 
changes is likely to be already in contradiction with observational
data. 

There are to our mind strong indications -- at least within the
realm of {\it Euclidean} quantum gravity -- that the topological sum
cannot be made meaningful, simply because {\it it results in too
many configurations contributing to the path integral}.
This is true even in dimension two, where toy models of quantum
gravity (in the form of generally covariant non-perturbative
Euclidean path integrals over geometries) can be defined and solved exactly.
In this case, no difficulty arises with the labelling of topologies, 
which amounts
to a single parameter $g$, the genus or number of handles (holes) of
the 2d geometry. The topological expansion in $g$ was the subject of
intense study in the early 1990s, because it is an example of the
non-perturbative sum over worldsheets of a bosonic string,
in a zero-dimensional target space. The problem in making
the sum well-defined stems from the factorial growth in $V$ of the 
number of inequivalent 2d surfaces of a given volume $V$.
Moreover, the coefficients in the $g$-expansion are positive,
obstructing Borel-summability, and no way has been found to 
define the non-perturbative sum unambiguously (see
\cite{fgz,adj} for detailed discussions and references). 

Given the recent successes in obtaining quantum gravity theories
from state sums over {\it Lorentzian} geometries in dimensions
two \cite{al,aal1,aal2,krist1,ackl,krist2} and three \cite{ajl2,ajlv,hexagon,ajl5}
(see also \cite{loll} for a recent set of lecture notes), the question
arises of how a topological sum can be incorporated in these models and
whether any progress can be made in performing the sum.
We have shown in \cite{lw} that for quantum gravity in two space-time
dimensions the problem is indeed ameliorated by going to a
Lorentzian signature: consideration of
their causal properties leads
to a natural restriction on the topology-changing geometries 
entering the regularized path integral, as will be explained below. 
The combined sum over
topologies and geometries can be performed exactly, and
possesses a well-defined double-scaling limit, involving both
the cosmological and the gravitational coupling constants,
$\Lambda$ and $G$. For $G\rightarrow 0$, standard Lorentzian
quantum gravity without holes is recovered, whereas for values
larger than zero, the presence of holes leads to an observable and 
non-local scattering of light rays traversing the space-time. At $G=G_{max}$,
the system undergoes a transition to a phase of ``handle condensation".
In addition to a further instance of how Lorentzian-ness and
causality lead to path integrals that are better behaved than their
Euclidean counterparts, this opens up a new playground for 
gravity-inspired 2d statistical models. -- The remainder of this
contribution describes the construction and solution of 
2d Lorentzian gravity with holes obtained from the discrete
one-step propagator \cite{lw}. A more detailed analysis 
is contained in a forthcoming publication \cite{lwnext}.

\section{Implementing topology change}\label{bad}

There is some freedom in how to include topology-changing 
1+1 dimensional space-time configurations in the gravitational
path integral. Our implementation will be minimal in the sense
that each hole will be allowed to 
exist for an infinitesimal proper-time interval
only. In our discrete, triangulated framework this will mean that
a hole will come into existence at some integer time $t$ and
disappear again at $t+1$. The number of allowed holes per
time step $\Delta t=1$ (in the continuum limit) will be arbitrary.
As in Lorentzian quantum gravity for fixed topology, all 
configurations possess a globally defined proper time variable. For
the sake of definiteness, we will work with spatially compact
slices. Therefore, by construction a spatial slice at some integer $t$
will have the topology $S^1$ of a circle, whereas for all times
in the open interval $]t,t+1[$ it will be split into a constant number
$g+1$ of $S^1$-components.

Although this seems the very mildest form of topology change 
imaginable in two dimensions, we will see that generic 
space-times of this kind are extremely ill-behaved in their
geometric and causal properties, even if there is only a single
hole in the entire space-time. The essential difference with the
Euclidean case is that the presence (almost everywhere) 
of a Lorentzian structure
allows us to quantify how badly causality is violated (as it
necessarily must be in a topology-changing geometry).
We will then argue for a restriction of the state sum to 
geometries whose causality violations are relatively mild.
This is motivated by the search for continuum limits which
do {\it not necessarily} exhibit macroscopic acausal and
therefore physically unacceptable behaviour (adopting a
similar line of argument as one would in 4d).

What we find is 
an exactly soluble model of 2d gravity with dynamical topology,
with a well-defined double-scaling
limit involving both the gravitational and cosmological coupling 
constants. Its acausal properties can be probed by light rays,
and get larger with increasing (renormalized) gravitational
constant $G$. This means that for $G\not= 0$ there is always a 
non-trivial effect coming from the ``infinitesimal" holes, which 
may still be compatible with observation if the measuring
instruments that could detect the acausality were not sufficiently
sensitive. However, as we will see, for sufficiently large $G$
any experimental detection threshold will eventually be
exceeded.
We interpret this behaviour as an a posteriori 
justification for having restricted the allowed space-time histories
in the first place, in the sense that it seems unlikely that a model
with significantly more general types of holes 
will possess any physically acceptable ground state whatsoever.
\begin{figure}[tbp]
\centerline{\scalebox{0.5}{\rotatebox{0}
{\includegraphics{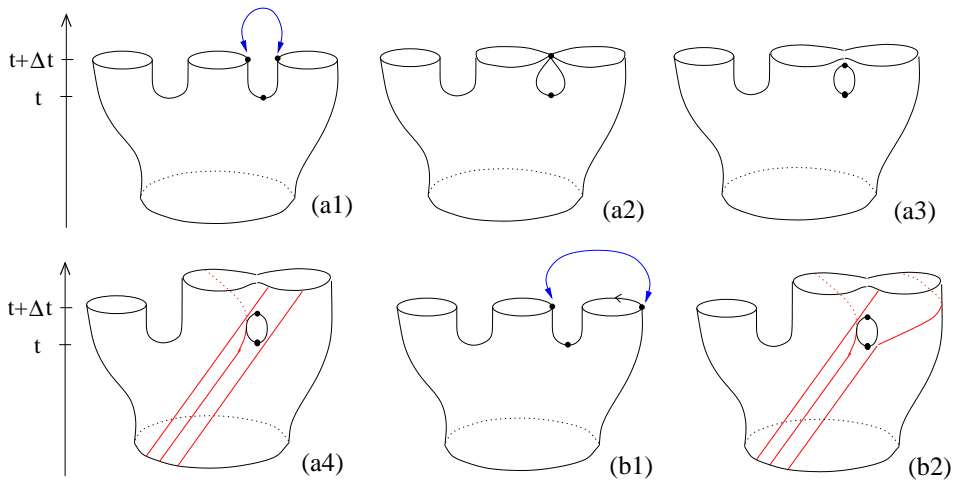}}}}
\caption[tubes]{A connected spatial slice splits into three components at time
$t$, (a1). In the first example, two circles at time $t+\Delta t$ are merged into 
one by identifying two points
which are timelike separated from the original branching point, (a2) and (a3).
Parallel light rays passing between $t$ and $t+\Delta t$ are unaffected unless 
they are scattered non-locally by the hole to another part of the manifold, 
as is the central light ray in (a4). -- If one of the merging points is
spacelike separated from the branching point at $t$, a twist (indicated by 
the arrow in the embedded picture, (b1)) is required before the regluing.
In the resulting geometry, the distance between two parallel light 
rays which pass on either side of the hole has jumped discontinuously
after the merger, (b2).}
\vspace{0.5cm}
\label{tubes}
\end{figure}

We will first give a qualitative description of space-time geometries
with ``bad" and ``not-so-bad" topology changes, and then present a
concrete realization of the latter within the framework of 2d Lorentzian
dynamical triangulations. The generation of holes of either type is
illustrated in Fig.\ref{tubes}. At time $t$,
an initial spatial slice $S^1$ splits into $g+1$ components, (a1), giving
rise to $g$ saddle points. The components evolve in time until 
$t+\Delta t$, where they re-unite to a single $S^1$. 
A difference now arises, depending on which pairs of points are
identified in the process of merging. In the not-so-bad topology change, the
upper saddle point of a hole is by definition 
time- or lightlike related to the
lower saddle point of the same hole, in either component, as indicated
in (a2), (a3) (for simplicity of illustration, we perform the merger only
for two of the components). A merger which is not of this type is
illustrated in Fig.\ref{tubes}, (b1). The marked point at time $t+\Delta t$
on the right-most cylinder component is supposed to lie outside the 
light cone of the lower saddle point\footnote{This is a slight abuse of
language, because there is of course no light cone right {\it at} the 
saddle point. The geometry of the neighbourhoods of such saddle 
points plays an important part in the analysis of
the causal properties of the associated space-times in the continuum 
\cite{dowker1,dowker2,dowker3,dowker4}.}.

To illustrate the qualitative difference between the two cases, we
follow a set of parallel light rays through the resulting space-times,
as indicated in Fig.\ref{tubes}, (a4) and (b2)\footnote{To keep the 
argument simple, we do not represent
any effects of the (generically non-vanishing) curvature on the light 
propagation, which is of a quite different (namely, local) nature.}.
In both cases, a light ray which ``hits the hole" is
scattered non-trivially to a different part of the manifold. However, in
the case where an additional relative twist of at least one of the cylinder
components is present (in Fig.\ref{tubes}, (b1) and (b2), the right cylinder has
been twisted by an angle $\pi$), 
there is another non-local effect,
consisting in a permutation of different finite sections of the 
propagating light front, which will persist after the hole has
disappeared. The effect on the two outer parallel light rays depicted in 
Fig.\ref{tubes}, (b2), is that they are still parallel after time $t+\Delta t$,
but their mutual distance will have jumped. 

Note that while the effect of the
direct scattering by a single hole will vanish in the limit as $\Delta t
\rightarrow 0$ (corresponding to the continuum limit in the discretized
model)\footnote{This
does not imply that the presence of more than one hole, coming from
different time slices, cannot
lead to observable effects. Indeed, this is exactly what we will find
below in the analysis of the continuum limit of the model.}, 
the effect of globally rearranging parts of space-time with
respect to each other for the ``bad" topology changes will persist in the 
same limit, and represents an observable, macroscopic violation of causality.
We will discard such configurations from the path integral, since
we do not think that these large-scale causality violations can
cancel out in any superposition of such geometries. Moreover,
they completely outnumber the geometries with ``not-so-bad"
topology changes. The precise definition of the resulting 2d quantum
gravity model, its continuum limit, and its physical properties will be the
subject of the following section.

\begin{figure}[tbp]
\centerline{\scalebox{0.6}{\rotatebox{0}
{\includegraphics{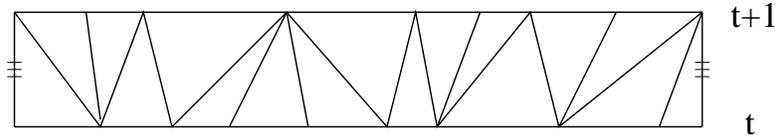}}}}
\caption[onestrip]{A strip $[t,t+1]$ of a 2d Lorentzian triangulated
space-time. The ends of the strip should be identified as indicated,
leading to a compact spatial geometry $S^1$.}
\label{onestrip}
\end{figure}

\section{Lorentzian quantum gravity with holes}

We will now discuss how to implement ``baby holes" of the type
introduced in the previous section explicitly
in the framework of piecewise linear two-manifolds. Recall that 
space-time geometries in Lorentzian dynamical triangulations are
constructed by gluing together strips of height $\Delta t=1$,
where $t$ is a discrete analogue of proper time. 
A given strip between integer times $t$ and $t+1$
consists of $N_t$ Minkowskian triangles (each with one spacelike 
and two timelike edges), 
and is periodically identified in the spatial direction, as illustrated
in Fig.\ref{onestrip}. 

\begin{figure}[tbp]
\centerline{\scalebox{0.6}{\rotatebox{0}
{\includegraphics{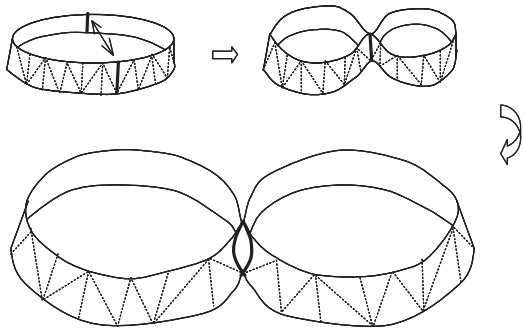}}}}
\caption[makinghole]{Constructing a strip with one hole by identifying two
of the timelike edges between times $t$ and $t+1$ of a regular Lorentzian
strip and separating them perpendicularly as indicated, thereby creating 
a hole between the two integer times.}
\label{makinghole}
\end{figure}

We create a hole of minimal time extension $\Delta t=1$, and 
associated with a ``not-so-bad" topology change, by identifying two 
timelike links in the same strip $[t,t+1]$ (these are links interpolating 
between the slices of constant time $t$ and $t+1$), and cutting
them open in the perpendicular direction, thereby creating two
cylinders and a minimal hole in between (Fig.\ref{makinghole}).
This process generates two
curvature singularities, at the beginning and end of the hole, which
after the Wick rotation will be of the standard conical type, and we
will choose their Boltzmann weights accordingly. As anticipated
earlier, the number of possible strip geometries of this type as a 
function of the total strip volume scales exponentially, and both 
the state sum and its continuum limit are completely well defined.

\begin{figure}[tbp]
\centerline{\scalebox{0.6}{\rotatebox{0}
{\includegraphics{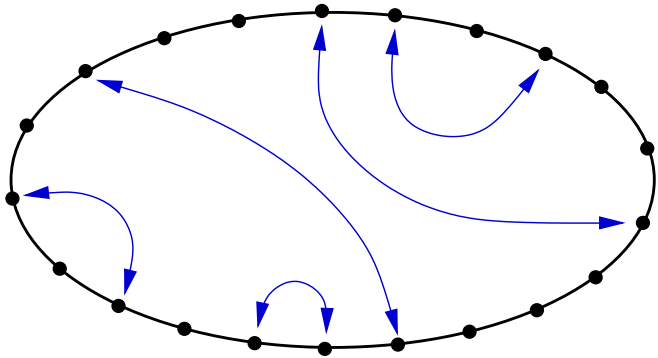}}}}
\caption[arch]{A typical graph showing the pairwise identification of 10 out of $N=21$
points, giving rise to a strip with 5 holes.}
\vspace{0.5cm}
\label{arch}
\end{figure}

As in the original Lorentzian model \cite{al}, it suffices to examine
the combinatorics of a single strip to
determine the bulk behaviour of the model in the continuum
limit (as well as the associated quantum Hamiltonian \cite{lwnext}). 
After the Wick rotation, the relevant partition function is 
\begin{equation}
Z(\lambda,\kappa)=\sum_{l_{in}}\sum_{l_{out}} G_{\lambda,\kappa}
(l_{in},l_{out};t=1), 
\label{part}
\end{equation}
where we have performed a sum over both the initial and final
boundary geometries of length $l_{in}$ and $l_{out}$, 
and where the propagator $G_{\lambda,\kappa}$
is given by
\begin{equation}
G_{\lambda,\kappa}(l_{in},l_{out};t=1)={\rm e}^{-\lambda (l_{in}+l_{out})}
\sum_{T|_{l_{in},l_{out}} }{\rm e}^{-\kappa g(T)}.
\label{prop}
\end{equation}
The sum in (\ref{prop}) 
is taken over all triangulated strip geometries with boundary
lengths $l_{in}$ and $l_{out}$, and $g\geq 0$ denotes the number of holes
of the strip. In writing eq.\ (\ref{prop}) we have used that
the discrete volume of a strip is given by $N\equiv N_t=l_{in}+l_{out}$, as in
Lorentzian gravity without topology change. Fixing $l_{in}$ and $l_{out}$
(and for convenience putting a mark on the entrance loop), the number 
\begin{equation}
\tilde G(l_{in},l_{out})=\biggl({l_{in}+l_{out}-1\atop l_{in}-1}\biggr)
\label{puregrav}
\end{equation}
of distinct (marked) interpolating strip triangulations without holes gives
rise to an overall factor $2^{N-1}$. 
For a given triangulated strip of volume $N=l_{in}+l_{out}$, holes 
are created according to the prescription given above and 
as shown in
Fig.\ref{makinghole}.
\begin{figure}[tbp]
\centerline{\scalebox{0.6}{\rotatebox{0}
{\includegraphics{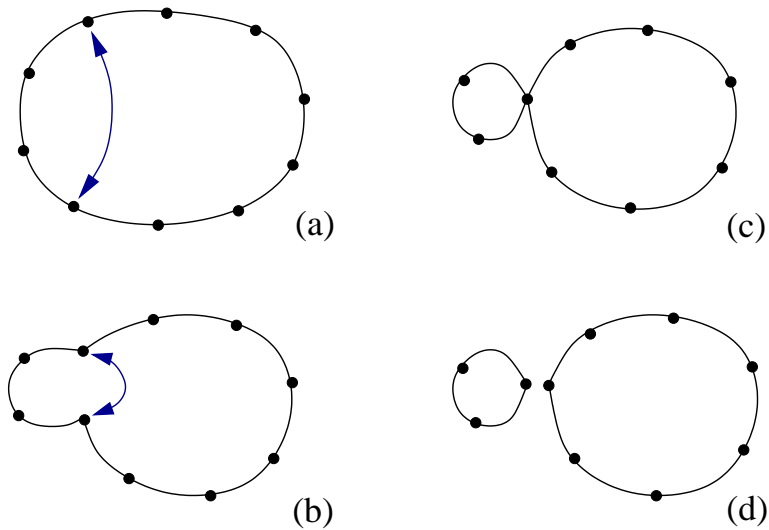}}}}
\caption[onehole]{Inserting a single hole into a regular strip of length
$N=10$ (a),
as it appears in a slice of constant time half-way between the
two boundaries. The strip is pinched along a pair of timelike links (appearing
as dots), as 
indicated by the arrows (b), until a figure eight is obtained (c), after which
the strip is separated into two cylinders as indicated in (d).}
\vspace{0.5cm}
\label{onehole}
\end{figure}
An alternative, planar representation of the creation of a single hole
is given in Fig. \ref{onehole},
which shows a cut through the strip halfway between times $t$ and $t+1$.
The $N$ timelike links of the strip appear as dots on the circle 
(Fig.\ref{onehole}a). The procedure 
for several holes is completely analogous. The only restriction one needs
to impose in order to obtain a well-defined two-geometry with $g+1$
cylindrical components is that the $g$ arrows identifying pairs of points in
the corresponding planar diagram should not cross each other.\footnote{The
{\it planarity} of  ``arrow diagrams" like Fig.\ref{arch} is responsible for an
exponential, as opposed to a factorial counting.} Also, we
will impose the regularity condition that there should be at most one arrow 
per point. This avoids some double counting of identical geometries and 
eliminates a few geometries with cylinders of the size of the cutoff, and
is not expected to affect the continuum limit in any way. Note also that
we are including some geometries where one or more cylinders 
degenerate to a point either at time $t$ or $t+1$; this is merely to 
simplify some of the combinatorial formulas and again will not have any
consequences for the continuum limit. 
Our next task is therefore to count the number of ways of inserting
$g$ holes into a strip of volume $N$, which is equivalent to counting
graphs with $N$ points and $g$ arrows, an example of which is
shown in Fig.\ref{arch}. This is readily done by noting that the number
of ways to pick $2g$ out of $N$ points, $2g\leq N$, is given by 
\begin{equation}
\biggl({N\atop 2g}\biggr),
\label{count1}
\end{equation}
since the $N$ points can be regarded as distinguishable (at large volumes
$N$, a generic triangulated strip without holes will not have any symmetries). 
For a given set of $2g$ points, we then have to count the number of ways
of connecting them by non-intersecting arches. Fortunately, this is a well-known 
combinatorial problem whose resolution is given by the so-called Catalan numbers
\begin{equation}
A(2g)=\frac{(2g)!}{g! (g+1)!}.
\label{catalan}
\end{equation}
The complete formula for the partition function (\ref{part}) is therefore
\begin{equation}
Z(\lambda,\kappa)= \frac{1}{2} \sum_{N=0}^\infty\ \sum_{g=0}^{[N/2]}\  \biggl({N\atop 2g}\biggr)
\ \frac{(2g)!}{g! (g+1)!}\  {\rm e}^{-2\kappa g} {\rm e}^{-(\lambda -\log 2)N}.
\label{z1}
\end{equation}
After an exchange of the two sums, both of them can be performed explicitly, leading
to
\begin{equation}
Z(\lambda,\kappa)=\frac{1}{2 (1-{\rm e}^{-(\lambda-\log 2)})}\ \frac{1-\sqrt{1-4 z}}{2z},
\label{partz}
\end{equation}
where the second term on the right-hand side\footnote{This term is
recognized as the generating function for the Catalan numbers, and has
previously appeared in a statistical model of certain ``decorated"
2d Lorentzian geometries without topology changes \cite{krist2}.}
depends only on the specific combination
\begin{equation}
z:={\rm e}^{-2\kappa}({\rm e}^{\lambda -\log 2}-1)^{-2}
\label{both}
\end{equation}
of the bare coupling constants $\kappa$ and $\lambda$. The partition function
(\ref{z1}) is convergent for (real) $\lambda >\log 2$ and $z<1/4$.
We are now interested in constructing a
continuum limit of $Z$. This will necessarily involve an infinite-volume limit 
$N\rightarrow\infty$.
It is straightforward to compute the expectation value of the discrete volume,
\begin{equation}
\langle N\rangle =-\frac{1}{Z}\frac{\partial Z}{\partial \lambda}=
\frac{ {\rm e}^{\lambda} } { ({\rm e}^\lambda-2) \sqrt{1-4z}}-1,
\label{nav}
\end{equation}
from which we deduce that the infinite-volume limit can be obtained
by letting $\lambda$ approach $\log 2$ from above, just like in standard
2d Lorentzian gravity. However, this is only consistent if one stays inside
the combined region of convergence of both $\lambda$ {\it and} $z$.
From the explicit form (\ref{both}) of $z$ this is only possible if one
scales the bare inverse gravitational coupling $\kappa$ in such a way
as to counterbalance the divergence coming from the inverse powers of
$({\rm e}^{\lambda -\log 2}-1)$. More specifically, if we make a standard
ansatz of canonical scaling for the cosmological coupling constant,
\begin{equation}
\lambda =\lambda^{crit}+a^2\Lambda +O(a^3)\equiv
 \log 2 +a^2\Lambda +O(a^3),
\label{cosren}
\end{equation}
where $\Lambda$ denotes the renormalized, dimensionful cosmological
constant, we obtain for any fixed value $z=c<1/4$ of $z$ an equation for
$\kappa$, namely,
\begin{equation}
\kappa =-\frac{1}{2} \log \biggl( c\, (a^2\Lambda)^2\biggr) +O(a),
\label{kapeq}
\end{equation}
which determines the leading-order behaviour of $\kappa$ as a
function of the cutoff $a$.
This relation can now be read as the defining equation for the renormalized
inverse gravitational coupling $\rm K$,
\begin{equation}
{\rm K} =\kappa -2 \log \frac{1}{a\sqrt{\Lambda}} +O(a),\;\;\;{\rm with}\;\;\;\;
{\rm K}:=\frac{1}{2} \log \frac{1}{c}.
\label{kapren}
\end{equation}
The logarithmic subtraction is what one would expect for
the renormalization of a dimensionless coupling constant.
Introducing the renormalized Newton's constant $G=1/$K, and substituting the 
expansions 
(\ref{cosren}) and (\ref{kapren}) into (\ref{partz}), one obtains to lowest order 
in $a$ 
\begin{equation}
Z=\frac{1}{a^2}\ \frac{ {\rm e}^{2/G}}{4\Lambda } 
\biggl( 1-\sqrt{1-4\, {\rm e}^{-2/G}}
\biggr) =: \frac{1}{a^2}\ Z^R(\Lambda,G),
\label{zdone}
\end{equation}
where we have performed a wave function renormalization to arrive at
a finite renormalized partition function $Z^R(\Lambda,G)$. In summary,
we have been led to (\ref{zdone}) by taking a well-defined
{\it double-scaling limit} of both the gravitational and the
cosmological coupling constants, very similar to what has always
been hoped for in 2d {\it Euclidean} quantum gravity (see, for example,
\cite{adj}, p.186ff). 

\begin{figure}[tbp]
\centerline{\scalebox{0.9}{\rotatebox{0}
{\includegraphics{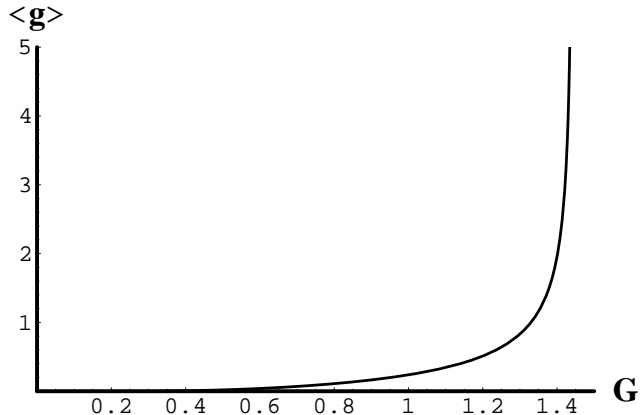}}}}
\caption[genplot]{The average number $\langle g\rangle$ of holes as a function
of the renormalized Newton constant $G$. }
\vspace{0.5cm}
\label{genplot}
\end{figure}

The physical properties of the resulting continuum theory are governed 
by the values of these two couplings. The expectation value of the space-time 
volume $V:=a^2 N$, computed in analogy with (\ref{nav}), is given
by the inverse cosmological constant,
\begin{equation}
\langle V\rangle =-\frac{1}{Z^R} \frac{\partial Z^R}{\partial\Lambda}=
\frac{1}{\Lambda},
\label{volume}
\end{equation}
as one may have anticipated. The role of the gravitational constant is
exhibited by computing the expectation value of the number of holes per
time interval,
\begin{equation}
\langle g\rangle=\frac{1}{2}\, G^2\frac{1}{Z^R}\frac{\partial Z^R}{\partial G}=
-\frac{1}{2}\biggl( 1-\frac{1}{\sqrt{1-4 {\rm e}^{-2/G}}}\biggr).
\label{holes}
\end{equation}
This function is plotted in Fig.\ref{genplot} in the range $G\in[0,2/\log 4]$.
For small coupling $G\approx 0$, there are basically no holes, up to
$G\approx 1.33$, where there is an average of a single hole in the
entire strip.
Beyond this value, the number of holes increases rapidly and diverges 
at the boundary $G=2/\log 4$ of the allowed interval.
The average genus is an interesting quantity because it relates in a
direct way to an ``observable",
namely, the part of a light beam that undergoes scattering when passing
through a Lorentzian space-time with baby holes. In first approximation,
this is given by
\begin{equation}
{\rm scattered\; portion\; of\; light\; beam}\; \propto \langle g\rangle \frac{T}{L},
\label{light}
\end{equation}
where $L$ is the characteristic linear spatial extension of the ``quantum
universe", and $T$  is the (continuum) time of propagation of the light beam.

\section{Conclusions}

We have formulated a regularized path integral over two-dimensional 
Lorentzian space-times, including a sum over topologies, and
shown that it possesses a non-trivial continuum limit. The Lorentzian
structure of the individual discretized geometries was used to
classify topology changes as ``bad" and ``not-so-bad" in terms of
their causal properties. The former were excluded from the sum over
space-times because they implied large-scale causality violations,
even for isolated holes of minimal duration. At the same time, this
prescription turned out to eliminate a factorial divergence in the
entropy part of the state sum, which in the past has proved troublesome in 
analogous, purely Euclidean gravitational path integrals.

The sum over the remaining topology-changing geometries can be
performed explicitly, and converges for suitable choices of the two bare
coupling constants. The infinite-volume is obtained by tuning the
cosmological constant to its critical value. In order not to run out
of the region of convergence of the wick-rotated partition function,
a simultaneous scaling of the gravitational coupling is required.
This double-scaling limit leads to an unambiguously defined 
non-perturbative theory of 2d quantum gravity, which one may
think of as a concrete realization of space-time foam in two
dimensions. Its physical properties
depend on the values of the two renormalized couplings, where the
cosmological constant $\Lambda$ sets the overall scale of the universe 
and the renormalized Newton constant $G$ determines the abundance
of microscopic holes.

\vspace{1cm}
\noindent {\it Acknowledgements.} WW thanks the organizers of the workshop
for a very pleasant meeting held in a wonderful location. Support through the
EU network on ``Discrete Random Geometry'', grant HPRN-CT-1999-00161,
is gratefully acknowledged.

\end{document}